\documentclass[aps,preprint,amsmath,amssymb,amsfonts,superscriptaddress]{revtex4-1}
\usepackage{color}
\usepackage{graphicx}   
\usepackage{array}
\usepackage{dcolumn}    
\usepackage{bm}         
\usepackage[normalem]{ulem}

\begin{document}
\title{Dynamic Anapole in Metasurfaces made of Sculptured Cylinders}

\author{Evangelia Takou}\email{etakou@physics.uoc.gr}
\affiliation{Institute of Electronic Structure and Laser, FORTH, 70013, Heraklion, Crete, Greece}
\affiliation{Department of Physics, University of Crete, 70013, Heraklion, Crete, Greece}
\author{Anna C. Tasolamprou}
\affiliation{Institute of Electronic Structure and Laser, FORTH, 70013, Heraklion, Crete, Greece}
\author{Odysseas Tsilipakos}
\affiliation{Institute of Electronic Structure and Laser, FORTH, 70013, Heraklion, Crete, Greece}

\author{Eleftherios N. Economou}
\affiliation{Institute of Electronic Structure and Laser, FORTH, 70013, Heraklion, Crete, Greece}
\affiliation{Department of Physics, University of Crete, 70013, Heraklion, Crete, Greece}

%

\date{\today}
\begin{abstract}
We present all-dielectric polaritonic metasurfaces consisting of properly sculptured cylinders to sustain the dynamic anapole, i.e. a non-radiating alternating current distribution. One way for the anapole to emerge, is by combining modes based on the first and the fourth Mie resonance of a cylinder made of high permittivity LiTaO$_3$ for operation in the low THz. The circular cross-section of each cylinder varies periodically along its length in a binary way, from small to large, while its overall circular symmetry has to be broken in order to remove parasitic magnetic modes. Small cross-sections are the main source of the \textit{electric dipole} Mie mode, while large cross-sections sustain the fourth, \textit{mixed toroidal dipole} Mie mode. With proper adjustment, the generated electric and toroidal moments interfere destructively producing a non-radiating source, the dynamic anapole, the existence of which is attested by a sharp dip in the reflection from the metasurface, due exclusively to electric and toroidal dipoles. Moreover, we show that, by breaking the circular symmetry of each cylinder, a substantial toroidal dipole emerges from the \textit{magnetic quadrupole} Mie mode, which in combination with the electric dipole also produces the dynamic anapole. The sensitivity of the anapole states to the material dissipation losses is examined leading to the conclusion that the proposed metasurfaces offer a scheme for realistically implementing the anapole.  
\end{abstract}

\maketitle


\section{\label{sec:Intro}Introduction}

Non-radiating sources have attracted strong interest both from the fundamental point of view of electrodynamics and for the applications of non-scattering objects \cite{Devaney19731044,Kim1986}. A type of non-radiating source that attracts increasing attention in the photonic community is the dynamic anapole, which is usually created by the superposition of electric and toroidal dipole moments \cite{Koshelev2019,Baryshnikova2019}. The toroidal dipole is a peculiar excitation that differs from the familiar electric and magnetic dipoles; the latter involve back-and-forth currents and closed circulation of currents respectively; the pure toroidal dipole rises from the poloidal currents circulating on a surface of a torus along its meridians. In spite of these quite different current distributions, both the electric and the toroidal dipoles emit radiation with the same angular momentum and parity properties, rendering their radiation pattern indistinguishable for any distant observer \cite{Afanasiev1995}; this is one of the reasons that toroidal multipoles are not considered in classical electrodynamics textbooks \cite{Dubovik1990,Radescu2002}; it is also the reason that electric and toroidal dipoles appear to be ideal for destructive interference and, hence, for the realization of the dynamic anapole, although it is quite difficult to match their strengths and isolate them from adjacent contributions. The evolution of metamaterials, artificial materials with engineered macroscopic electromagnetic properties, has provided the tool for overcoming this last difficulty \cite{Kaelberer1510} and for the realization of the dynamic anapole, the nontrivial, non-radiating source which is the objective of the present work.

As it was mentioned before, the dynamic anapole can be created by a toroidal dipole (\textbf{T}) oscillating out of phase relative to the electric dipole (\textbf{p}) leading thus to destructive interference of the radiated field in the far-field zone. The relation that guarantees this interference reads:
\begin{equation} \label{eq:1}
\textbf{p}+ik\textbf{T}=0
\end{equation}
where $k=\omega/c$. Although in the far-field region the fields vanish, this is not the case for the source region; this is due to the fact that the difference of the vector potentials \textbf{A}$_\mathrm{T}$(\textbf{r},$t$) and \textbf{A}$_\mathrm{p}$(\textbf{r},$t$) of a toroidal and an electric dipole emission, respectively, cannot be eliminated via a gauge transformation. More explicitly, if one considers point-like sources (as there is no limitation upon the size of the source) of electric and toroidal dipoles superimposed under the relation of Eq.~(\ref{eq:1}), the corresponding vector potential $\mathrm{\Delta}\textbf{A}=\textbf{A}_\mathrm{T}-\textbf{A}_\mathrm{p}$ reads:
\begin{equation}
\Delta\textbf{A}=\nabla \left( \textbf{T} \cdot \nabla \left( \frac{e^{ikr}}{r} \right) \right) + 4\pi \delta^3(\textbf{r})\cdot \textbf{T}
\end{equation}

The second term in the expression for the net vector potential is not-trivial, since it does not vanish upon application of the curl. In other words, this indicates that the net vector potential cannot be eliminated at all points in space in any gauge \cite{Zheludev2016}. In the static case, $k = 0$, the electric dipole moment disappears, \textbf{p} = 0, making the static anapole synonymous with the static toroidal dipole. It has been argued that the point dynamic anapole may be viewed as the basic building block out of which an arbitrary non-radiating source can be composed of \cite{Nemkov2017}. 

The first anapole state was verified experimentally in the GHz spectrum using metallic split ring resonators \cite{Fedotov2013}. Simpler geometries followed and the first experimental observation of the dynamic anapole in the visible, in a stand alone dielectric particle (Si nanodisk), was reported in \cite{Miroshnichenko2015}, as well as in structures of core-shell nanowires \cite{Liu:15} and in spheres \cite{Wei:16}. Hybrid situations of coexisting magnetic and electric anapole states have also been investigated for the case of a high index spherical particle \cite{Kuznetsov2017}. The non-radiating response of the anapole modes accompanied by the enhanced near fields has found numerous applications such as in cloaking \cite{Liu2017}, harmonic generation \cite{Grinblat2016,Grinblat2017,Grinblat2017_19}, nanoscale lasers \cite{ToteroGongora2017}, high \textit{Q} factor devices \cite{Liu:17}, sensing \cite{Ospanova2018,Beccherelli2018}, Raman scattering enhancement \cite{doi:10.1021/acsphotonics.8b00480} and many more. The far field extinction combined with the maximized electromagnetic energy concentration within the nanostructures has also opened the possibility of intensity enhancement in all-dielectric nanostructures \cite{Miroshnichenko2015,Zenin2017,doi:10.1021/acsphotonics.7b01440}, which was accomplished by utilizing Si nanodisks (slotted in the latter case). 

Most of the works discussing the anapole excitation in metamaterials involve the manipulation of the toroidal moment through various structural schemes. Metallic metamaterials with asymmetric inclusion, like asymmetric metallic bars \cite{article} or properly adjusted in U-shaped metaatoms \cite{doi:10.1002/adma.201606298} have been proven to provide control over the toroidal excitation. Also in the case of all-dielectric metamaterials control over the toroidal excitation has been provided by the shape, size and formation of the nanoparticles, for example the size of silicon disks  \cite{Miroshnichenko2015,Liu:17}, the formation of silicon fins \cite{31old_29new}, the number, shape and formation of infinite length polaritonic rods \cite{32old_30new,33old_31new,34old_32new,PhysRevX.5.011036,38old_34new} etc. Remarkably, it has been also recently realized that toroidal dipole moment exists within the natural TE$_{01}$ mode (fourth mode in ascending frequency) of an infinite length single cylinder \cite{Liu:15,33old_31new,8107810}. The toroidal dipole moment in this mode is accompanied by a strong electric dipole moment due to the asymmetric positive and negative polarization currents. 

In this paper, we present a way to manipulate this \textit{mixed toroidal dipole} Mie mode by combining it with an additional \textit{electric dipole moment} in a polaritonic metasurface. Beginning from the simple infinite, uniform cylinder, made of LiTaO$_3$, we design a sculptured cylinder with alternating large and small cross-sections that sustain the long sought pure dynamic anapole based on the combination of TE$_{00}$ and TE$_{01}$ type Mie modes. Additionally, we show that the anapole state can also occur by the combination of the electric dipole Mie mode TE$_{00}$ and a mixed mode of toroidal dipole character that emerges from a hybrid magnetic quadrupole Mie mode by breaking the circular symmetry of the cylinder. The signature of the anapole is recorded in the power reflected by the metasurface. In particular, a clear, sharp reflection dip appears that comes almost exclusively from the contributions of the electric and toroidal dipoles; this is proven by the fact that the contributions of the electric and toroidal dipole moments reproduce the total reflected field from the metasurface. At the anapole state, our scheme provides sufficiently increased toroidal moment contribution to cancel out the electric dipole one. We evaluate also the role of material dissipation losses concluding that the designed metasurface provides a scheme for realistically  implementing the anapole. 

The manuscript is organized as follows: In Sec.~\ref{sec:second}, we present the modes sustained by an infinite and uniform cylinder and characterize their electromagnetic character with a focus on the electric and mixed toroidal dipole character. These modes are properly modified by our design of a sculptured cylinder (with the binary distribution of its cross-section) serving as the building block for the proposed metasurface. In Sec.~\ref{sec:third} we present the simulation results of the reflection power by the metasurface designed to sustain the anapole state due to the mixed toroidal and the electric dipole modes. In Sec.~\ref{sec:fourth} we discuss the structure's sensitivity to the dissipation losses. Finally, in Sec.~\ref{sec:fifth} we present the design principles, implementation, and dissipation losses sensitivity of a modified metasurface based on a toroidal component emerging from quadrupole-like Mie mode and electric dipole. Then we conclude our work. 

\section{\label{sec:second} Mixed electric and toroidal modes of a cylinder and principles of design}

We investigate the possibility of a dynamic anapole by considering a metasurface consisting of dielectric cylinders; the circular cross-section of each of them varies periodically along its length in a binary way as shown in Fig.~\ref{fig:rods}(c). The values of the small and the large cross-sections as well as their corresponding lengths are to be fixed in such a way as to satisfy the anapole condition of Eq.~(\ref{eq:1}), as accurately as possible. We begin the study by focusing on the Mie eigenmodes of an isolated polaritonic cylinder, infinite and uniform along its length coinciding with the $\hat{\textbf{z}}$-axis direction [Fig.~\ref{fig:rods}(a)]. We assume $E_\mathrm{z}$ polarization and eigensolutions whose wavevector remains strictly in the \textit{xy} plane, $k_\mathrm{z}=0$. Initially we consider that the radius of the infinite LiTaO$_3$ rod is equal to $R_\mathrm{c}=15$~$\mu$m. We have chosen the LiTaO$_3$ polaritonic material to drastically reduce the ohmic losses while maintaining high permittivity. In a polaritonic material the induced polarization currents are due to the AC motions of the cations and the anions; their dielectric function has the form: 
\begin{figure*} [!htbp]
\includegraphics[scale=0.85]{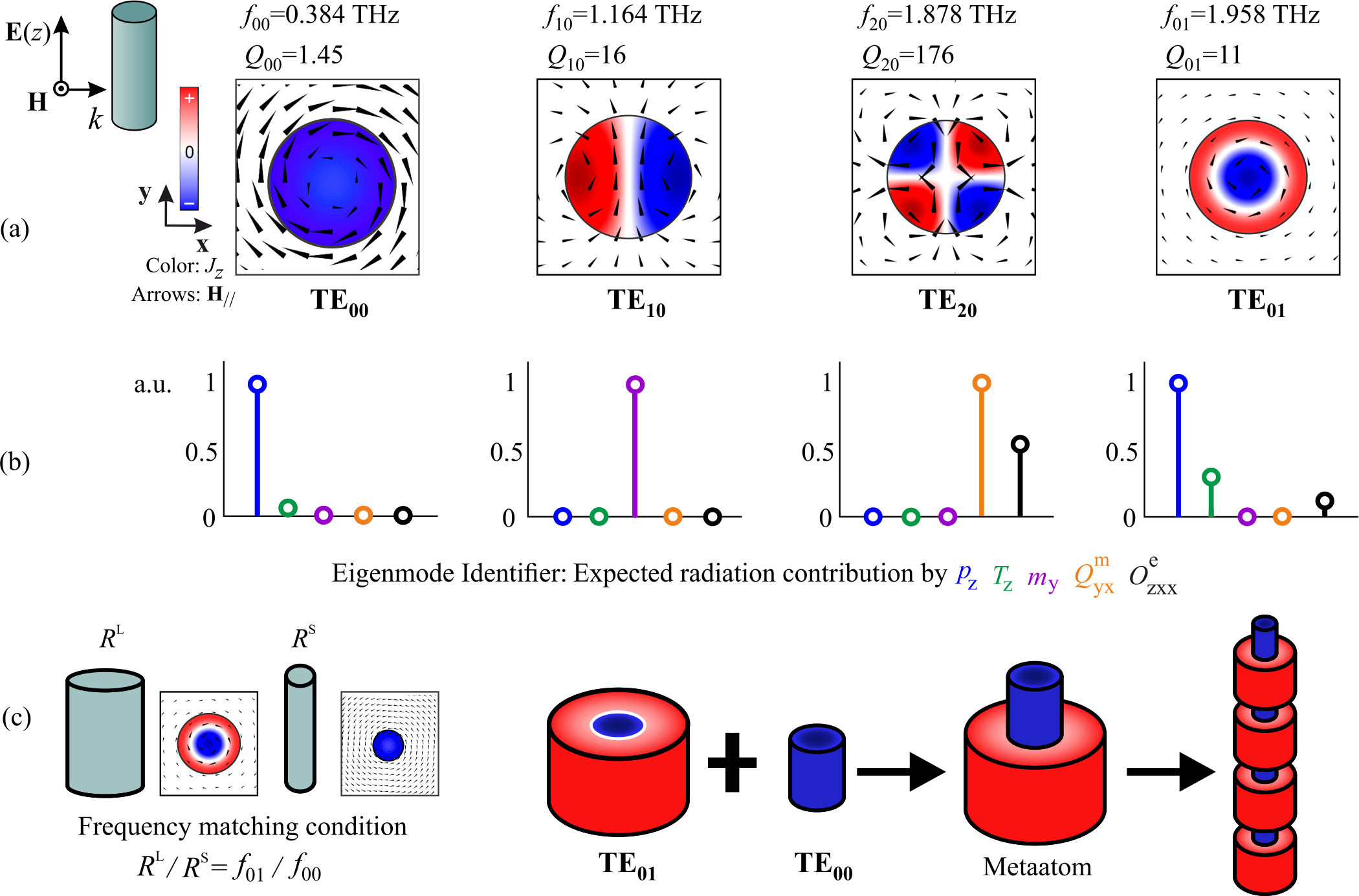}
\caption{\label{fig:rods} (a) First four transverse electric eigenmodes of a single infinite length and uniform circular cross-section polaritonic rod: polarization current distribution $J_\mathrm{z}$ (color), magnetic field lines (arrows), resonant frequency and quality factor are shown. (b) Eigenmode identifier regarding the contributions coming from the electric dipole moment $p_\mathrm{z}$, the toroidal dipole moment $T_\mathrm{z}$, the magnetic dipole moment $m_\mathrm{y}$, the magnetic quadrupole moment $Q^{\mathrm{m}}_{\mathrm{yx}}$ and the electric octupole moment $O^{\mathrm{e}}_{\mathrm{zxx}}$; these radiation contributions are normalized in terms of the dominant one in each mode individually. The character of the shown modes is: TE$_{00}$, electric dipole; TE$_{10}$, magnetic dipole; TE$_{20}$, magnetic quadrupole; and TE$_{01}$, \textit{mixed} toroidal dipole; the term \textit{mixed} is adopted since both electric and toroidal dipoles participate in the eignemode, in spite the fact that the magnetic field lines suggest a predominantly toroidal moment. (c) The unit cell of each cylinder (the so called metaatom) consists of two cylindrical pieces one of  smaller and the other of larger cross-section such that the TE$_{00}$ eigenfrequency of the first to coincide with the TE$_{01}$ eigenfrequency of the second, opening thus the path for the realization of the dynamic anapole.}
\end{figure*}
\begin{align} 
\epsilon(\omega) = \epsilon_1+i\epsilon_2 &=\epsilon_{\infty} \frac{\omega^2_\mathrm{L}-\omega^2-i\omega \gamma}{\omega^2_\mathrm{T}-\omega^2-i\omega \gamma} \\ 
\epsilon(0) \equiv \epsilon_0 &=\epsilon_\infty \left( \frac{\omega_\mathrm{L}}{\omega_\mathrm{T}} \right)^2
\end{align}

For LiTaO$_3$ the transverse optical resonance frequency is $\omega_{\mathrm{T}} /2 \pi =26.7$~THz, the longitudinal resonance frequency is $\omega_\mathrm{L} /2 \pi =46.9$~THz, $\epsilon_\infty=13.4$ and $\gamma /2 \pi = 0.94$~THz, with the static permittivity equal to $\epsilon_0=41.34$. In the low THz regime, the dielectric function has a real constant value of $\epsilon_1 \sim  41$, while the dissipation losses are in the order of few $10^{-3}$. The dissipation losses are low enough to be considered negligible for our initial investigation; extensive study of the structures sensitivity to losses is provided in Sects.~\ref{sec:fourth} and \ref{sec:fifth}.

Figure~\ref{fig:rods}(a) presents the $E_\mathrm{z}$ polarized Mie eigenmodes in ascending frequency, the polarization current distribution, the resonant frequency and the quality factor, sustained by an isolated, free-space standing, LiTaO$_3$ cylinder of radius $R_\mathrm{c}=15$~$\mu$m and permittivity $\epsilon_1=41$. The sustainable modes are calculated by eigenvalue analysis. Figure~ \ref{fig:rods}(b) presents an electromagnetic character identification of each mode which is attained by calculating the dominant expected radiation contributions (amplitude) by the multipole moments in each eigenmode. Multipole moments are vectorial functions of the spatial distribution of polarization currents with expressions that incorporate the toroidal multipoles found in Refs.~\cite{PhysRevB.89.205112,38old_34new} (Appendix C of Ref.~\cite{PhysRevB.89.205112}). Essentially, they describe the electromagnetic character (electric, magnetic, toroidal) of the local sources that are created by the circulation of alternating currents. In this case, the only nonzero component of the polarization current is $J_\mathrm{z}$ and the nonzero multipole radiation contributions come from the electric dipole moment $p_\mathrm{z}$, the toroidal dipole moment $T_\mathrm{z}$, the magnetic dipole moment $m_\mathrm{y}$, the magnetic quadrupole moment $Q^\mathrm{m}_\mathrm{yx}$, and the electric octupole moment $O^\mathrm{e}_{\mathrm{zxx}}$. There is also an additional contribution coming from the electric quadrupole moment $Q^\mathrm{e}_{\mathrm{xz}}$, which in the two-dimensional space coincides with the magnetic dipole moment $m_\mathrm{y}$ and is therefore omitted from Fig.~\ref{fig:rods}(b). Note also that the vector components depend on the rotational symmetry of the modes and the reference cartesian unit system as defined in Fig.~ \ref{fig:rods}(a). They are both selected to match the metasurface set up of Sec.~\ref{sec:third}. In Fig.~\ref{fig:rods}(b), in each individual eigenmode the expected radiation contributions by the multipoles are normalized in terms of the dominant contribution. As seen in Figs. \ref{fig:rods}(a) and \ref{fig:rods}(b) the first mode, TE$_{00}$ (subscript denoting zero radial and zero azimuthal variation), corresponds to the electric dipole and exhibits a very low quality factor $Q_{00}\sim 1.5$ which means that, when excited, is expected to strongly radiate in a broad frequency range. It is characterized by a strong electric dipole moment contribution from $p_\mathrm{z}$ and a residual toroidal dipole moment contribution from $T_\mathrm{z}$. The next mode, TE$_{10}$ (subscript denoting one azimuthal and zero radial variation), is the magnetic dipole with $Q_{10}=16$ and is characterized by a strong contribution coming from $m_\mathrm{y}$ alone. Next in frequency is the TE$_{20}$ magnetic quadrupole with a distinct magnetic quadrupole moment signature and a high quality factor $Q_{20}\sim 176$; in this case the dominant expected contribution comes from $Q^\mathrm{m}_{\mathrm{yx}}$, but at the same time there is a strong contribution that comes from $O^\mathrm{e}_{\mathrm{zxx}}$. The fourth mode is TE$_{01}$, with a zero azimuthal and one radial variation and a low quality factor $Q_{01}\sim 11$ indicative of its radiative nature. As seen in Fig.~\ref{fig:rods}(b), the radiation contribution is expected to come mainly from the electric dipole $p_\mathrm{z}$ and secondarily from the toroidal dipole $T_\mathrm{z}$; therefore it is denoted as mixed toroidal dipole in spite the fact that the magnetic lines suggest a predominantly toroidal dipole moment.   

The idea behind achieving the anapole in this system is to match the eigenfrequencies of the mixed toroidal dipole mode TE$_{01}$ and the electric dipole mode TE$_{00}$. This can be achieved by properly adjusting the diameter of the two sections in the binary metaatom, as presented in Fig.~\ref{fig:rods}(c), in such a way that the radiation by the electric dipole $p_\mathrm{z}$ and the toroidal dipole $T_\mathrm{z}$ cancels out. The extra degree of freedom offered by the height of the two sections will be used in our goal to obtain the dynamic anapole. In the system discussed here, where the rod is isolated in free-space and only dielectric materials with very low or no dispersion are involved, the frequency varies directly with the size of the rods. Therefore, the matching of the TE$_{00}$ and TE$_{01}$ mode occurs in cylinders with ratio of the two cross-sectional radii approximately equal to the ratio of the TE$_{00}$ and TE$_{01}$ frequencies $R_\mathrm{S}/R_\mathrm{L}\sim$~1/5. The subscripts ‘$S$’, ‘$L$’ stand for the small and large cross-sections that sustain the TE$_{00}$ and TE$_{01}$, respectively. 

\section{\label{sec:third} The anapole as a combination of electric and toroidal dipole modes in the dielectric metasurface}

\begin{figure} [!htbp]
\includegraphics{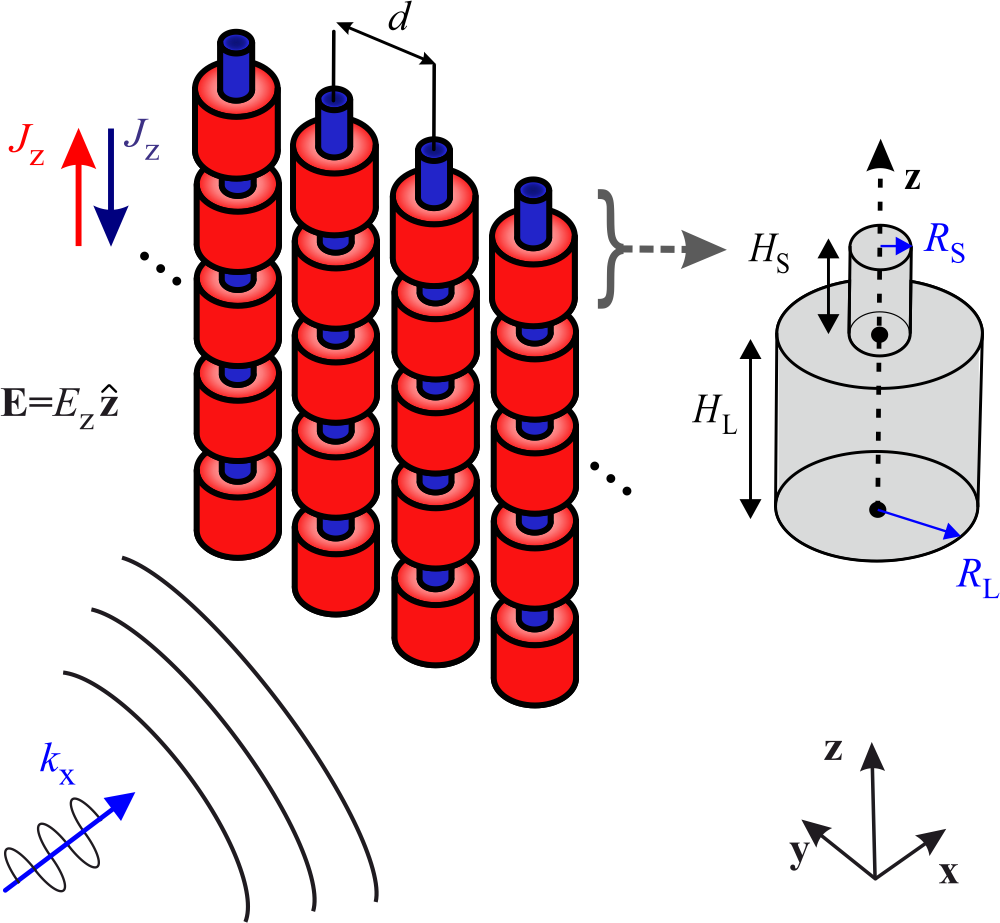}
\caption{\label{fig:metasurface} Schematic representation of a metasurface composed by periodically repeating along the $\hat{\textbf{y}}$-axis the sculptured cylinder shown in Fig.~\ref{fig:rods}(c). The incident wave is $\hat{\textbf{z}}$-polarized and normal incidence along the $\hat{\textbf{x}}$-axis is assumed. The parameters for the small and large cross-sections of the unit cell are radii, $R_\mathrm{S}$ and $R_\mathrm{L}$ and heights $H_\mathrm{S}$ and $H_\mathrm{L}$, respectively.}
\end{figure}

Using the sculptured cylinders with the periodic binary cross-section as building blocks, we construct the metasurface shown in Fig.~\ref{fig:metasurface}. 
The radius of the small cross-section and the large cross-section areas are $R_\mathrm{S}$, $R_\mathrm{L}$ respectively and the corresponding heights are $H_\mathrm{S}$ and $H_\mathrm{L}$. The period along the $\hat{\textbf{y}}$-axis is equal to $d=40$~$\mu$m. A linearly polarized (\textbf{E} parallel to \textbf{$\hat{\textbf{z}}$}-axis) plane wave impinges normally on this metasurface provoking the generation of currents that modify the scattered far-field as expressed in the transmission and reflection coefficients. Our goal is to achieve the dynamic anapole initially by manipulating the electric and mixed toroidal dipole moments; as far as the latter is concerned, we enhance its contribution by increasing the relevant volume of the larger cross-section [sustaining TE$_{01}$ as depicted in Fig.~\ref{fig:rods}(a) and \ref{fig:rods}(c)], while remaining in the subwavelength regime. In fact, if the ‘$S$’ and ‘$L$’ lengths are equal the electric dipole fully dominates the response; to remedy this, the TE$_{00}$-sustaining section has to be much shorter, compared to the TE$_{01}$-sustaining section.

\begin{figure} [!htbp]
\includegraphics{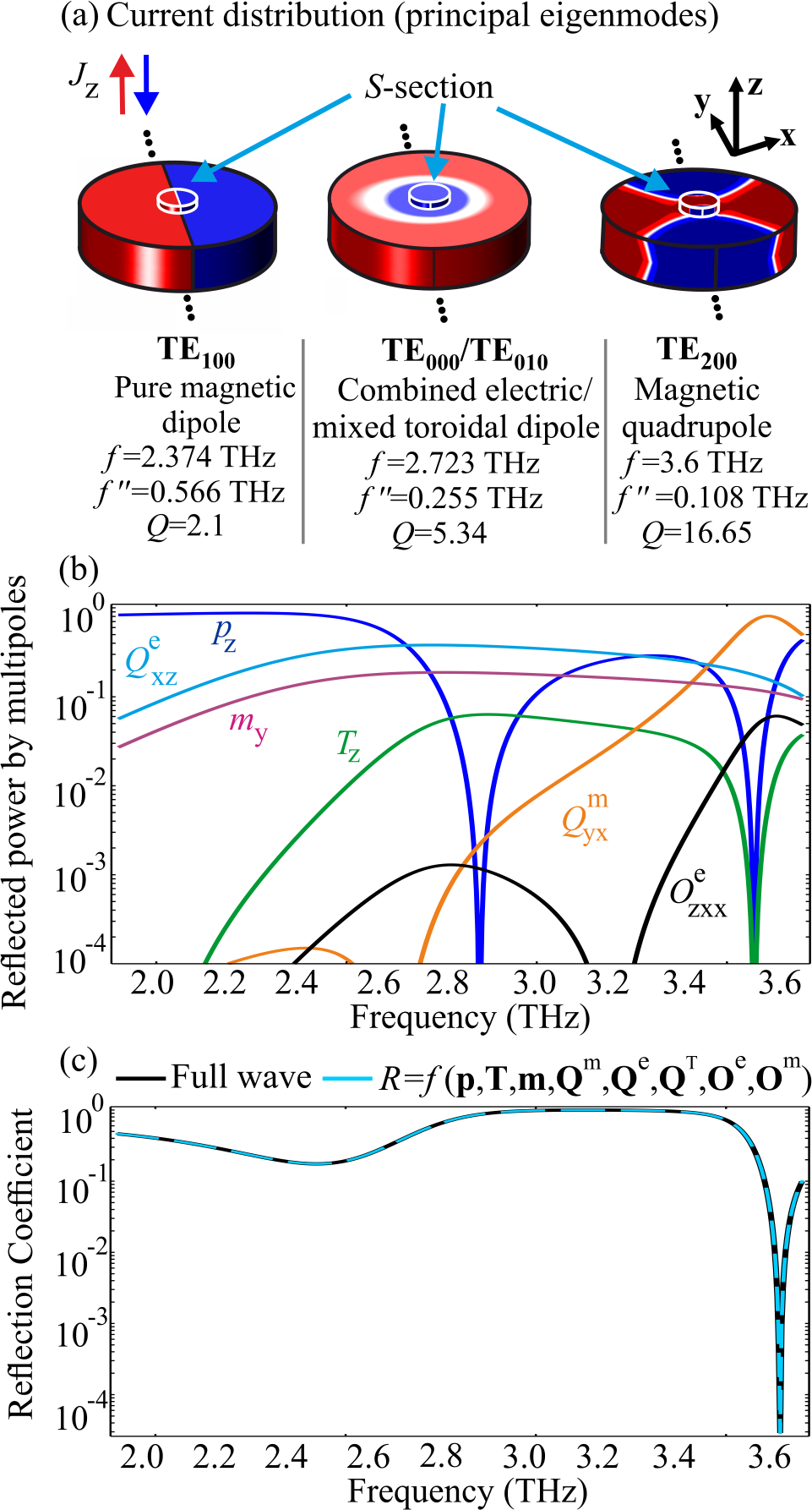}
\caption{\label{fig:nonoptimized} (a) Eigenfrequencies (real and imaginary parts), and polarization current distributions of the $k_\mathrm{y}=k_\mathrm{z}=0$, $\textbf{E}=E_\mathrm{z}\hat{\textbf{z}}$ eigenmodes in the metaatom part of the whole metasurface as shown in Fig.~\ref{fig:metasurface}; the shown eigenmodes are the pure magnetic dipole TE$_{100}$, the mixed toroidal TE$_{010}$ and the magnetic quadrupole TE$_{200}$. (b) Reflected power contributions by the individual moments; notice the significant strength and the almost coincidence of the contributions of the magnetic dipole $m_\mathrm{y}$ and electric quadrupole $Q^\mathrm{e}_{\mathrm{xz}}$. (c) Reflection coefficients for the frequency range 1.9-3.7~THz calculated by the full wave analysis and the summation of all the amplitude contributions. Notice that the reflection dip occurs close to the electric and the toroidal dipole dips, indicating that the magnetic dipole/quadrupole and the electric quadrupole/octupole cancel out at the frequency of the reflection dip.}
\end{figure}

We begin our investigation by assuming the non-optimized values of the design parameters for the metasurface: $H_\mathrm{L}=6$~$\mu$m, $H_\mathrm{S}=1$~$\mu$m, $R_\mathrm{L}=15$~$\mu$m, $R_\mathrm{S}=3$~$\mu$m and $d=40$~$\mu$m. The ratio between the large ‘$L$’ and small ‘$S$’ cross-section ensures the spectral matching for the electric and mixed toroidal dipole modes. It should be noted here that the periodicity along the $\hat{\textbf{y}}$-axis and the $\hat{\textbf{z}}$-axis corrugation is expected to change the landscape of the sustained eigenmodes, their properties and moments identification with respect to the isolated cylinder of Fig.~\ref{fig:rods}. However, certain aspects such as some relative spectral positions, the radiation trends and the current distributions of the electric, magnetic and toroidal modes still survive. In  Fig.~\ref{fig:nonoptimized}(a), the eigenfrequencies (both real, $f$, and imaginary parts, $f''$) of the $k_\mathrm{y}=k_\mathrm{z}=0$ eigenmodes of the whole metasurface, shown in Fig.~\ref{fig:metasurface}, are presented together with the current distributions of these eigenmodes within the metaatom part [see Fig.~\ref{fig:rods}(c)]; although the eigenfrequencies have been shifted considerably relative to the values of a single uniform cylinder [shown in Fig.~\ref{fig:rods}(a)], the current distributions are still, surprisingly, almost identical to those of a single uniform cylinder. In fact, we used the same names and the same notations for the eigenmodes of the metasurface as for those of a single uniform cylinder. Notice also that due to the electromagnetically small size of the cylinders height ($H_\mathrm{L}\sim \lambda_\mathrm{n}/2.5$, $H_\mathrm{S}\sim \lambda_\mathrm{n}/15$, where $\lambda_\mathrm{n}$ is the wavelength within the dielectric atom, $\lambda_\mathrm{n}=\lambda_0/n=15$~$\mu$m for $f=3$~THz), the corresponding modes have no considerable variation in the $\hat{\textbf{z}}$-axis direction and this is the reason that they are denoted as TE$_{\mathrm{mn0}}$ [i.e. the third subscript referring to the $\hat{\textbf{z}}$-axis (axial) variation is zero]; thus TE$_{100}$ is the notation for the magnetic dipole, TE$_{000}$/TE$_{010}$ for the mixed toroidal (or, more accurately, for the combined electric dipole and mixed toroidal), and TE$_{200}$ for the quadrupole. 

Figure~\ref{fig:nonoptimized}(b) presents the reflected power associated with the excited multipoles by a plane wave normally incident on the metasurface. The reflected power from each multipole is a function of the incident wave induced polarization currents at each frequency and adds up to the total reflected field \cite{PhysRevB.89.205112,38old_34new}. The multipole moments that contribute mainly to the scattered field are: the electric dipole moment, $p_\mathrm{z}$, the magnetic dipole moment $m_\mathrm{y}$, the toroidal dipole moment $T_\mathrm{z}$, the electric quadrupole $Q^\mathrm{e}_{\mathrm{xz}}$, the magnetic quadrupole $Q^{\mathrm{m}}_{\mathrm{yx}}$ and the electric octupole $O^\mathrm{e}_{\mathrm{zxx}}$ (in this $\hat{\textbf{z}}$-dependent metaatom, $Q^\mathrm{e}_{\mathrm{xz}}$ is not identical to $m_{\mathrm{y}}$ as in the  $\hat{\textbf{z}}$-independent case). There is also some weak contribution from $Q^\mathrm{T}_{\mathrm{xz}}$ and $O^\mathrm{m}_{\mathrm{yxx}}$ (not shown here). A surprise comes by comparing the eigenvalues of the metasurface shown in the data of Fig.~\ref{fig:nonoptimized}(a) with the data of Fig.~\ref{fig:nonoptimized}(b). A normally incident plane wave is expected to excite the eigenmodes of the metasurface possessing the same \textbf{k} character and hence local extrema in the reflected power contributions are expected to appear at these eigenfrequencies. Instead, this occurs only for the magnetic quadrupole excited mode which exhibits a prominent peak at its eigenfrequency at 3.6~THz; however, no peaks appear in Fig.~\ref{fig:nonoptimized}(b) either at the magnetic dipole eigenfrequency (2.374~THz) or at the mixed toroidal eigenfrequency (2.723~THz); a possible explanation of this unexpected feature can be attributed to the very large imaginary parts of the eigenfrequencies of these eigenmodes which apparently round-off the expected peaks. In Fig.~\ref{fig:nonoptimized}(c) the total reflection coefficient in the frequency range 1.9-3.7~THz has been calculated by two methods: (i) by full wave analysis and (ii) by the summation of the various multipole amplitudes that contribute to the reflected field. The agreement between the full wave analysis and the analytical multipole expansion formulation is excellent as shown in Fig.~\ref{fig:nonoptimized}(c) (validating thus the imperative presence of the toroidal dipole in the multipole expansion). 

Returning to the fact that the imaginary parts of the eigenfrequencies of the magnetic dipole and the mixed toroidal dipole are very large (implying that the corresponding $Q$ factors are very small), we stress that this holds to even larger degree for the electric dipole eigenmode associated with the larger section of the metaatom in spite of its much lower eigenfrequency. As a result of the consequent broadness of these eigenmodes, their contribution to the reflection is substantial even for frequencies far away from their eigenfrequencies. With that in mind, we may interpreter the results of Fig.~\ref{fig:nonoptimized}(b): There is a broad spectral region of magnetic dipole/electric quadrupole $m_\mathrm{y}$ and $Q^\mathrm{e}_{\mathrm{xz}}$ dominance which comes from the highly radiating magnetic dipole eigenmode with resonance at 2.374~THz, possessing the second lowest $Q$ factor (after the electric dipole). The spectral signature of the electric dipole contribution $p_\mathrm{z}$ is more complicated due to the low $Q$ factor of the electric dipole sustained in each of the two sections of the metaatom shown in Fig.~\ref{fig:rods}(c), as well as the electric dipole character that is found in many excited modes originating from the uneven distribution of upward and downward currents.  In Fig.~\ref{fig:nonoptimized}(b), we notice two electric dipole, $p_\mathrm{z}$, reflection dips. As mentioned, there is a large $p_\mathrm{z}$ tail coming from, the much lower in frequency TE$_{000}$ mode associated with the $L$ section of the unit cell which is very broad (due to its very low $Q$-factor) and thus overlaps with the frequency region of 2-3~THz. The combined TE$_{000}$/TE$_{010}$ has also at the $S$ section of the metaatom a pure electric dipole moment [and an electric dipole moment at the $L$ section, see Fig.~\ref{fig:rods}(b) associated with the TE$_{01}$ eigenmode]. All these electric dipole contributions coming from different eigenmodes suffer a destructive interference at frequencies around 2.85~THz and 3.55~THz. The first dip comes mainly from the electric dipole contribution of the combined TE$_{000}$/TE$_{010}$. The second dip in the electric dipole contribution occurs around the magnetic quadrupole excited mode, at 3.6~THz, and is expected to be due, besides the TE$_{200}$ and the combined TE$_{000}$/TE$_{010}$, to higher frequency excited modes as well. The signature of the magnetic quadrupole $Q^\mathrm{m}_{\mathrm{yx}}$ (accompanied by the electric octupole $O^\mathrm{e}_{\mathrm{zxx}}$) is more clear since it appears dominantly in the spectral vicinity of the TE$_{200}$ excited mode and exhibits a high $Q$ factor as seen from the sharp $Q^\mathrm{m}_{\mathrm{yx}}$ (and $O^\mathrm{e}_{\mathrm{zxx}}$) resonance at 3.6~THz.  

For the realization of the anapole we need to achieve the condition of Eq.~\ref{eq:1}. As shown in Fig.~\ref{fig:nonoptimized}(b) the toroidal dipole reflection power $T_\mathrm{z}$ exhibits a rather weak but decent presence in the spectrum under consideration. In particular, in the spectral neighborhood of the combined TE$_{000}$/TE$_{010}$, $T_\mathrm{z}$ is featureless. However, there is a $T_\mathrm{z}$ dip at 3.55~THz which indicates that there is a significant amount of toroidal dipole within the excited TE$_{200}$ quadrupole mode [the toroidal dipole stems from the modification of the Mie quadrupole mode of Fig.~\ref{fig:rods}(c) due to the $\hat{\textbf{z}}$-axis corrugation and the $\hat{\textbf{y}}$-axis periodicity] and interferes destructively with the $T_\mathrm{z}$ tail of TE$_{000}$/TE$_{010}$. On the other hand, Fig.~\ref{fig:nonoptimized}(b) seems to indicate that the sought after anapole could potentially be found where the reflected power from $p_\mathrm{z}$ and $T_\mathrm{z}$ cross (two positions around 2.85~THz). Nevertheless, the toroidal dipole moment is still significantly weaker than the electric one around this spectral region. A more serious obstacle is the dominance of the magnetic dipole and the electric quadrupole over the toroidal in the spectral region of interest as witnessed in Fig.~\ref{fig:nonoptimized}(b). Thus, in order to achieve the anapole we need to enhance the toroidal dipole moment and eliminate the competition coming from the magnetic multipole contributions. Up to now we exploited the different size of the cross sections. Other degrees of freedom are the heights $H_\mathrm{S}$ and $H_\mathrm{L}$ of the small $S$ and large $L$ parts of the metaatom. In fact, the resonant frequencies of the magnetic modes are expected, as a result of their complicated current distributions, to be much more sensitive to the heights than the uniform electric modes; this provides a strong design tool for its isolation. Thus, the strategy to strengthen the relative toroidal component is to assume a greater height $H_\mathrm{L}$. 

\begin{figure} [!htbp]
\includegraphics[scale=0.8]{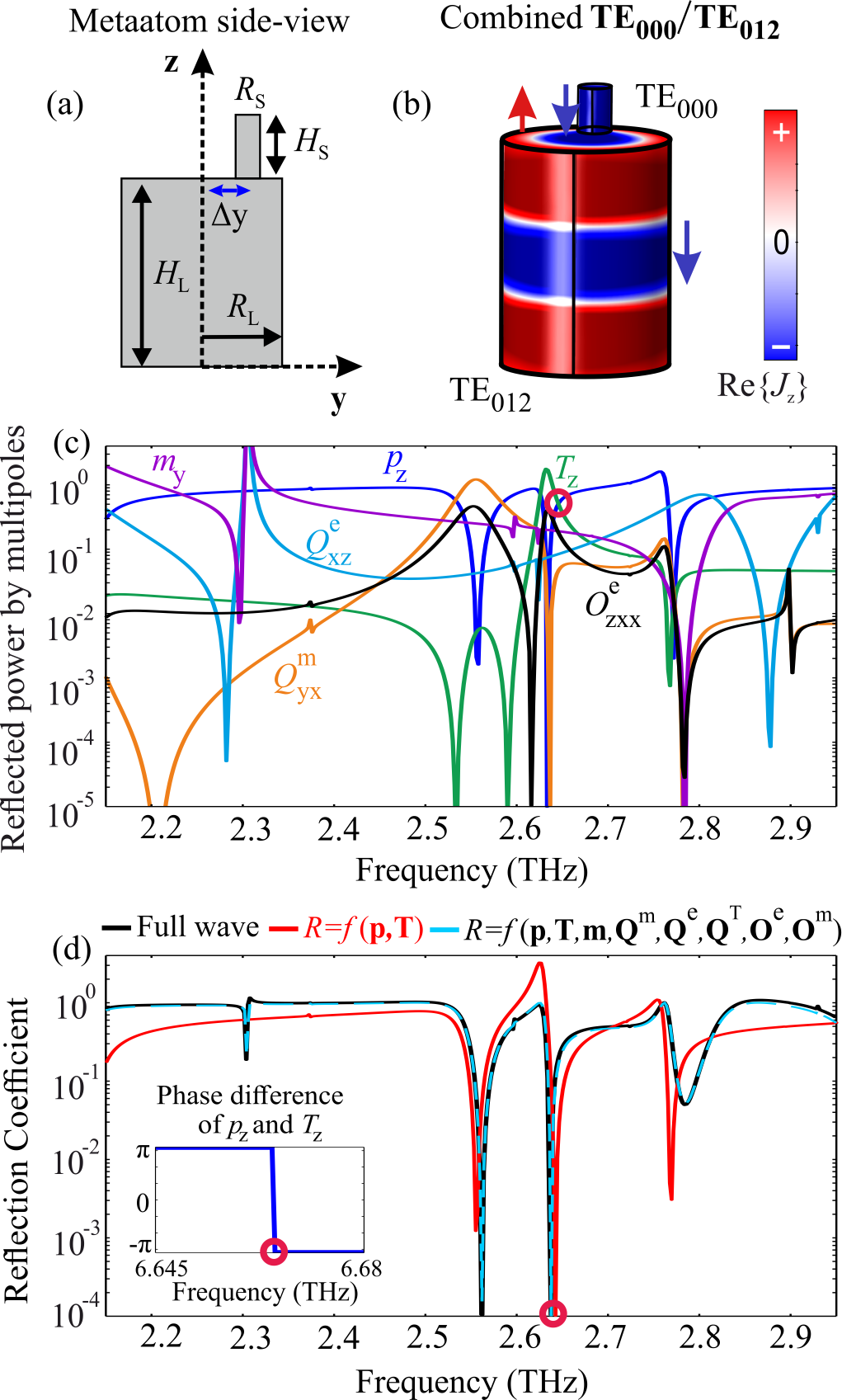}
\caption{\label{fig:config1_optimized} (a) Schematic representation (side-view) of the unit cell with a displaced smaller cylinder. Its center is placed at $y=10$~$\mu$m with respect to the axis of the larger cylinder. (b) Polarization current distribution of the TE$_{012}$ mode which is characterized by the reversal of the current flows within the unit cell along the $\hat{\textbf{z}}$-axis (the third subscript "2" denotes this reversal). The TE$_{012}$ mode is sustained by the larger cylinder comprising the unit cell, while the small cross section sustains the TE$_{000}$ mode. (c) Moments contribution to the reflected power, for the optimized metasurface with a unit cell as in Fig.~\ref{fig:config1_optimized}(a). The combined TE$_{000}$/TE$_{012}$ eigenmode of Fig.~\ref{fig:config1_optimized}(b) lies at 2.646~THz and exhibits a $T_\mathrm{z}$ power exceeding the $p_\mathrm{z}$ power. (d) Total reflection coefficient (directly from the full wave, black; and by adding all amplitudes, blue) and reflection coefficient including only the electric and toroidal dipole moments amplitudes (red) for the range 2.15-2.95~THz. The red curve is almost identical to the total reflection coefficient (a small shift is due to the magnetic moments contribution) showing that the sharp dip at 2.646~THz (i.e. the appearance of anapole behavior) is indeed mainly due to the destructive interference of electric and toroidal dipoles. This is further supported by phase difference $\mathrm{\Delta} \phi=-\pi$ between the $p_\mathrm{z}$ and the $T_\mathrm{z}$ dipoles (inset).}
\end{figure}

In this way, we can excite higher order modes in the $\hat{\textbf{z}}$-axis direction where the current is allowed to flow reversely within a single metaatom. In particular, we have found that if we set the parameters around the values $H_\mathrm{L}=40 $~$\mu$m, $H_\mathrm{S}=8$~$\mu$m and $R_\mathrm{L}=15$~$\mu$m and $R_\mathrm{S}=3$~$\mu$m for the heights and the cross-sectional radii respectively, while the periodicity is $d=35$~$\mu$m, we do get a relevant enhancement of the toroidal moment. At these heights, $H_\mathrm{S} \sim \lambda_\mathrm{n}/2$ and  $H_\mathrm{L} \sim 2.5 \lambda_\mathrm{n}$ (where $\lambda_n = \lambda_0 / n=15$~$\mu$m for the central frequency of 3~THz) a drastic modification of the combined TE$_{000}$/TE$_{012}$ mode occurs: while there is no $\hat{\textbf{z}}$-axis variation in the small cross-section part of the metaatom, a strong variation appears along the $\hat{\textbf{z}}$-axis for the large cross-section part as the third subscript "2" indicates. The distribution of the polarization currents for the combined TE$_{000}$/TE$_{012}$ is shown in Fig.~\ref{fig:config1_optimized}(b). Figure~\ref{fig:config1_optimized}(a) presents the schematic of the side-view of the metaatom. 
Notice that we have induced an asymmetric placement of the smaller cylinder along the $\hat{\textbf{y}}$-axis at a position $y=10$~$\mu$m. By introducing this asymmetry, we are able to further modify the magnetic excitations and in particular to suppress their contribution at the vicinity of the combined TE$_{000}$/TE$_{012}$ eigenmode (the electric and the toroidal modes are not affected to the same extent as the magnetic ones by this asymmetry as they do not possess azimuthal variations). Another way to tackle the magnetic quadrupole could be by assuming elliptical cross-section and increasing its spectral separation like in \cite{32old_30new}, but this did not provide an improved result. By enhancing the toroidal contribution and by eliminating the magnetic quadrupole we are able to satisfy the anapole condition of Eq.~(\ref{eq:1}). Towards this goal, we also numerically optimized the geometrical parameters of the structure as follows: $H_\mathrm{L}=39$~$\mu$m and $H_\mathrm{S}=7.8$~$\mu$m, $R_\mathrm{L}=14.625$~$\mu$m and $R_\mathrm{S}=2.95$~$\mu$m and $d=34.125$~$\mu$m. Figure~\ref{fig:config1_optimized}(c) presents the reflected (by the numerically optimized metasurface) power of the multipoles. The multipoles reflection contributions present more complicated yet as expected resonant features compared with those in Fig.~\ref{fig:nonoptimized}(b), as a result of the asymmetric positioning of the small cross-section cylinder. In Fig.~\ref{fig:config1_optimized}(d), the overall reflection \textit{R} by the metasurface is presented both by employing the direct full wave and by summing the amplitudes of all multipoles, $R=f(\textbf{p},\textbf{T},\textbf{m},\textbf{Q}^\mathrm{m},\textbf{Q}^\mathrm{e},\textbf{Q}^\mathrm{T}, \textbf{O}^\mathrm{m},\textbf{O}^\mathrm{e})$; if in the last sum we keep  only the amplitudes of the electric and the toroidal dipoles we obtain what is denoted by $R=f(\textbf{p},\textbf{T})$. The latter is compared with the overall reflection coefficient, leading to the conclusion that the zero reflection at the vicinity of the TE$_{000}$/TE$_{012}$ resonance ($f_{000/012}=2.646$~THz) is due to the destructive interference of the electric and the toroidal amplitudes; this conclusion is further supported by the fact that the toroidal and electric dipole moments meet with an opposite phase (e.g. $\mathrm{\Delta} \phi=-\pi$) at this frequency. A minor discrepancy in frequency in the order of 0.2$\%$ is observed and at the exact frequency of the anapole, a small radiation leakage of 10$\%$ coming from the reflection power contributions of all other multipoles is recorded. Thus, we have obtained beyond any doubt, the creation of an anapole by the destructive interference of electric and toroidal dipole amplitudes almost without interference of any other multipole moments. This achievement was realized by our optimized metasurface composed of sculptured cylinders periodically placed along the $\hat{\textbf{y}}$-axis; each cylinder is formed by asymmetric metaatoms [see Fig.~\ref{fig:config1_optimized}(a)] periodically repeated along the $\hat{\textbf{z}}$-axis. The circular symmetry breaking was essential in removing parasitic magnetic contributions to the anapole.
\section{\label{sec:fourth} Metasurface sensitivity to material losses}

\begin{figure*} [!htbp]
\includegraphics{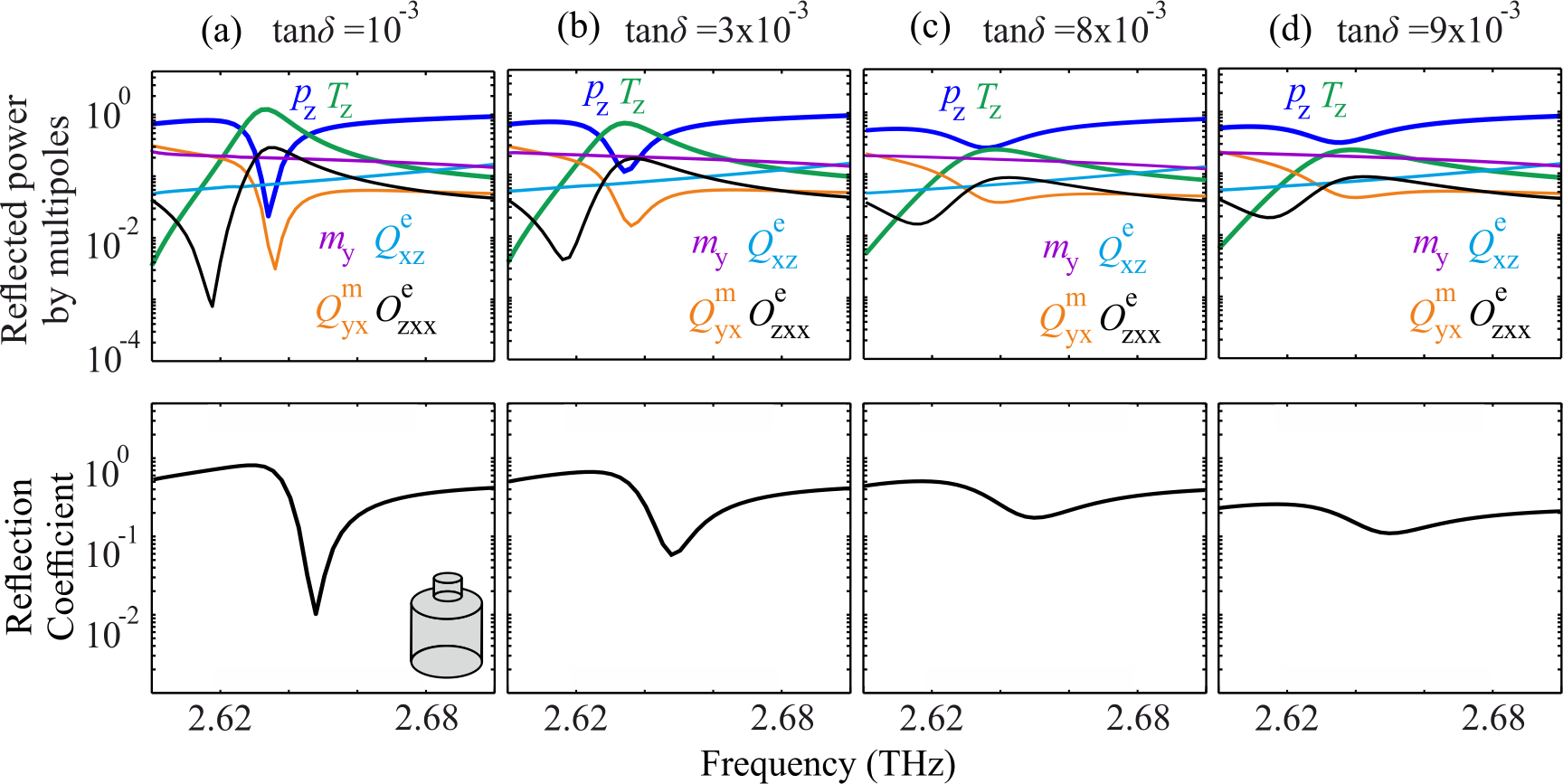}
\caption{\label{losses1} Moment contributions to the reflected power, for the optimized structure of Fig.~\ref{fig:config1_optimized}(a) (top panels) and total reflection (bottom panels) assuming material losses with (a) $\tan\delta=1\times 10^{-3}$, (b) $\tan\delta=3 \times 10^{-3}$, (c) $\tan\delta=8\times 10^{-3}$ and (d) $\tan\delta=9\times 10^{-3}$. Above $\tan\delta \sim 8\times 10^{-3}$ the electric dipole moment contribution increases extensively, it becomes dominant while the toroidal dipole moment contribution decreases and the crossings between the power reflected from $p_\mathrm{z}$ and $T_\mathrm{z}$ required for the anapole condition, seize to occur.}
\end{figure*}

Up to now, we have obtained the anapole excitation in the polaritonic metasurface assuming no material losses. In what follows, we examine the effect of the material losses on the anapole state. Notably, it appears that the anapole excitation in our system exhibits a considerable sensitivity to the material losses. Figure~\ref{losses1} presents the multipole contributions to the reflected power (top panels), as well as the overall reflection coefficient (bottom panels) for the optimized metasurface presented in Fig.~\ref{fig:config1_optimized}(a) assuming a material with variable losses.
The frequency range of the calculations is focused around the anapole state of the optimized structure (Fig.~\ref{fig:config1_optimized}) and is equal to 2.6-2.7~THz. For the real part of the material permittivity we use the nominal value for the polaritonic LiTaO$_3$, $\epsilon_1=41$, and various values of fictitious losses with tangent ranging from $\tan \delta=10^{-3}$ [Fig.~\ref{losses1}(a)] to $\tan \delta=9\times 10^{-3}$  [Fig.~\ref{losses1}(d)]. Low losses, with tangent in the order of few $10^{-3}$ have essentially little effect for the anapole realization, which is the case of the LiTaO$_3$ in the frequency range under consideration ($\tan\delta \sim 2.5\times 10^{-3}$). Of course, the resonance is not as sharp and as deep as in the absence of losses. This is to be expected, since in the anapole the fields at the metasurface are quite strong (in spite of their absence in the far zone) and consequently there are substantial material (not radiation) losses. This behavior is evident in Fig.~\ref{losses1}(a) and Fig.~\ref{losses1}(b), where the calculated response of the metasurface for material with $\tan\delta =10^{-3}$ and $\tan\delta =3\times 10^{-3}$ is shown, respectively. However, as losses increase, the crossing of the $p_\mathrm{z}$ and $T_\mathrm{z}$ dipoles becomes borderline at $\tan\delta =8\times 10^{-3}$ [Fig.~\ref{losses1}(c)] and beyond this value there is no crossing indicating that the condition required for the anapole is not satisfied anymore. As a result, the tendency of disappearance of the broad anapole dip is due to both material and radiation losses. In the regime beyond $\tan\delta =8\times 10^{-3}$ the electric dipole moment contribution becomes more dominant, while the toroidal dipole moment contribution diminishes. We note here that similar effect would be observed in a medium with presumable gain (not shown here), where an uneven increase of the electric dipole (dominant) and the toroidal dipole tends to eliminate the crossing of the reflected power from $p_\mathrm{z}$ and $T_\mathrm{z}$ and consequently the existence of anapole.
\newpage
\section{\label{sec:fifth} Anapole attained by hybrids of mixed electric and toroidal dipole modes }

Apart from the exploitation of the fundamental electric dipole and mixed toroidal dipole mode, the binary metaatom offers the possibility to explore an anapole excitation generated by the contribution of higher and hybrid modes that entail toroidal dipole moments. This could for example be explored in the spectral vicinity of the TE$_{200}$ resonance at 3.6~THz [Fig.~\ref{fig:nonoptimized}(a)]. This resonance includes a significant but hidden presence of toroidal dipole moment $T_\mathrm{z}$ simultaneously with the magnetic quadrupole $Q_{\mathrm{yx}}^\mathrm{m}$ and the other multipoles. By breaking the symmetry of the metaatom this hidden toroidal character is revealed and isolated from the quadrupole and other contributions offering an effective tool towards realization of the anapole. In Ref.~\cite{32old_30new}, it has been shown that this can be accomplished by introducing elliptical cross-sections in the metaatoms. In the present case, we further break the circular symmetry by introducing an additional perforation in the large cross-section of the metaatom of Fig.~\ref{fig:config1_optimized}. 

\begin{figure*} [!htbp]
\includegraphics[scale=0.85]{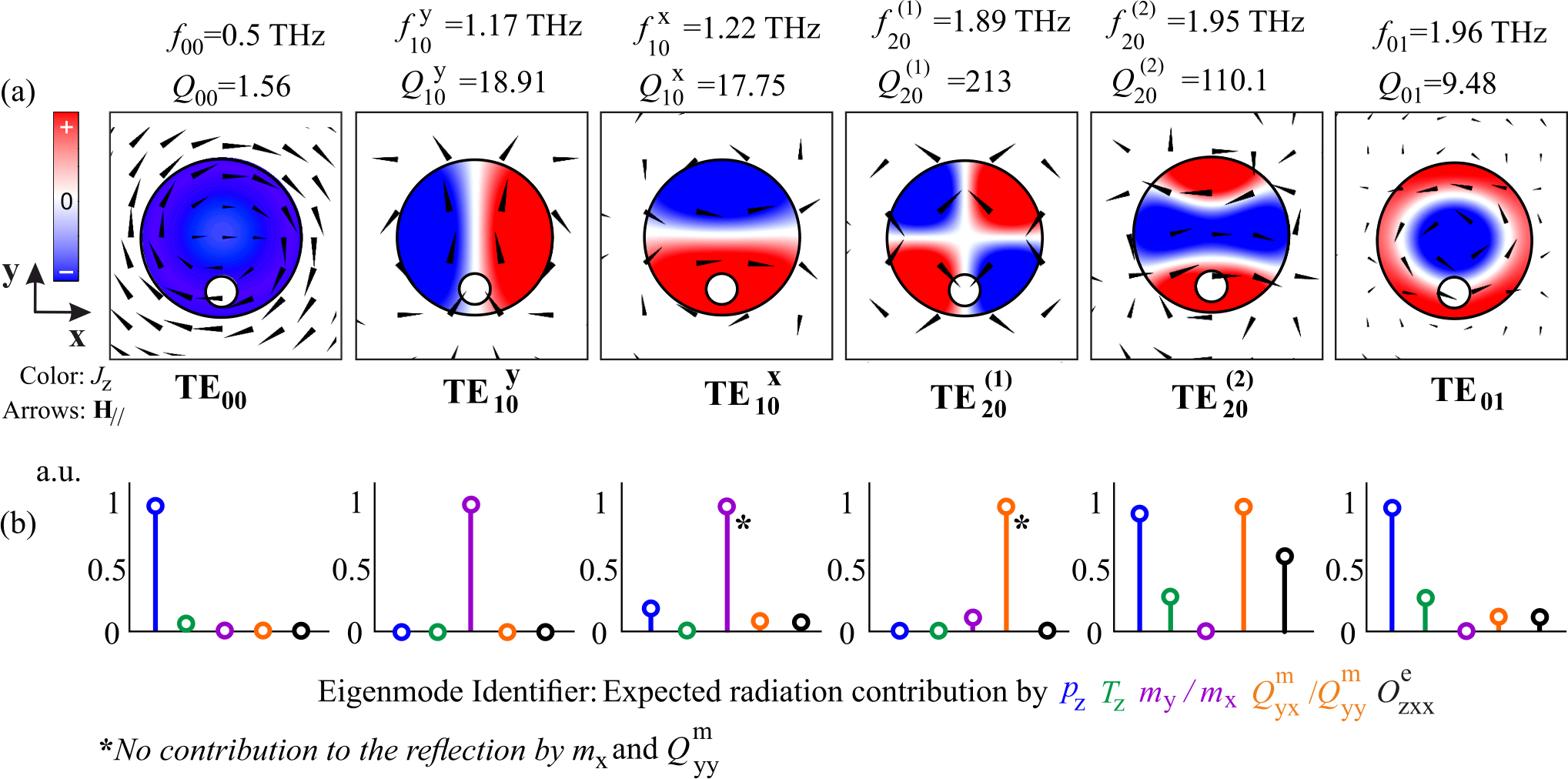}
\caption{\label{fig:perforated_eigen} (a) First six transverse electric eigenmodes of a single infinite length uniform polaritonic rod with a circular perforation; current distribution (color), resonant frequency, quality factor, and magnetic field lines (arrows) are shown. We assume $E_\mathrm{z}$ polarization and eigensolutions whose wavevector remains strictly in the $xy$ plane, $k_\mathrm{z}=0$. The radius of the cylinder is $R_\mathrm{c}=15$~$\mu$m and the radius or the perforation $r_\mathrm{p}=3$~$\mu$m placed at $y=-10$~$\mu$m. The sustained modes are the electric dipole, TE$_{00}$, two splitted magnetic dipoles, TE$_{10}^\mathrm{y}$ and TE$_{10}^\mathrm{x}$, two splitted magnetic quadrupoles TE$_{20}^{(1)}$ and TE$_{20}^{(2)}$ and the mixed electric and toroidal dipole, TE$_{01}$. Notice that the expected radiation corresponding to the TE$_{20}^{(2)}$ mode has been dynamically altered by the perforation compared to the TE$_{20}$ eigenmode of Fig.~\ref{fig:rods}(b). (b) Eigenmode identifier with respect to the expected radiation contributions coming from the electric dipole $p_\mathrm{z}$, the toroidal dipole $T_\mathrm{z}$, the magnetic dipole $m_\mathrm{y}$, magnetic quadrupole $Q^\mathrm{m}_{\mathrm{yx}}$ and the electric octupole $O^\mathrm{e}_{\mathrm{zxx}}$. The radiation power contributions are normalized with the dominant one in each mode individually.}
\end{figure*}

\begin{figure} [!htbp]
\includegraphics{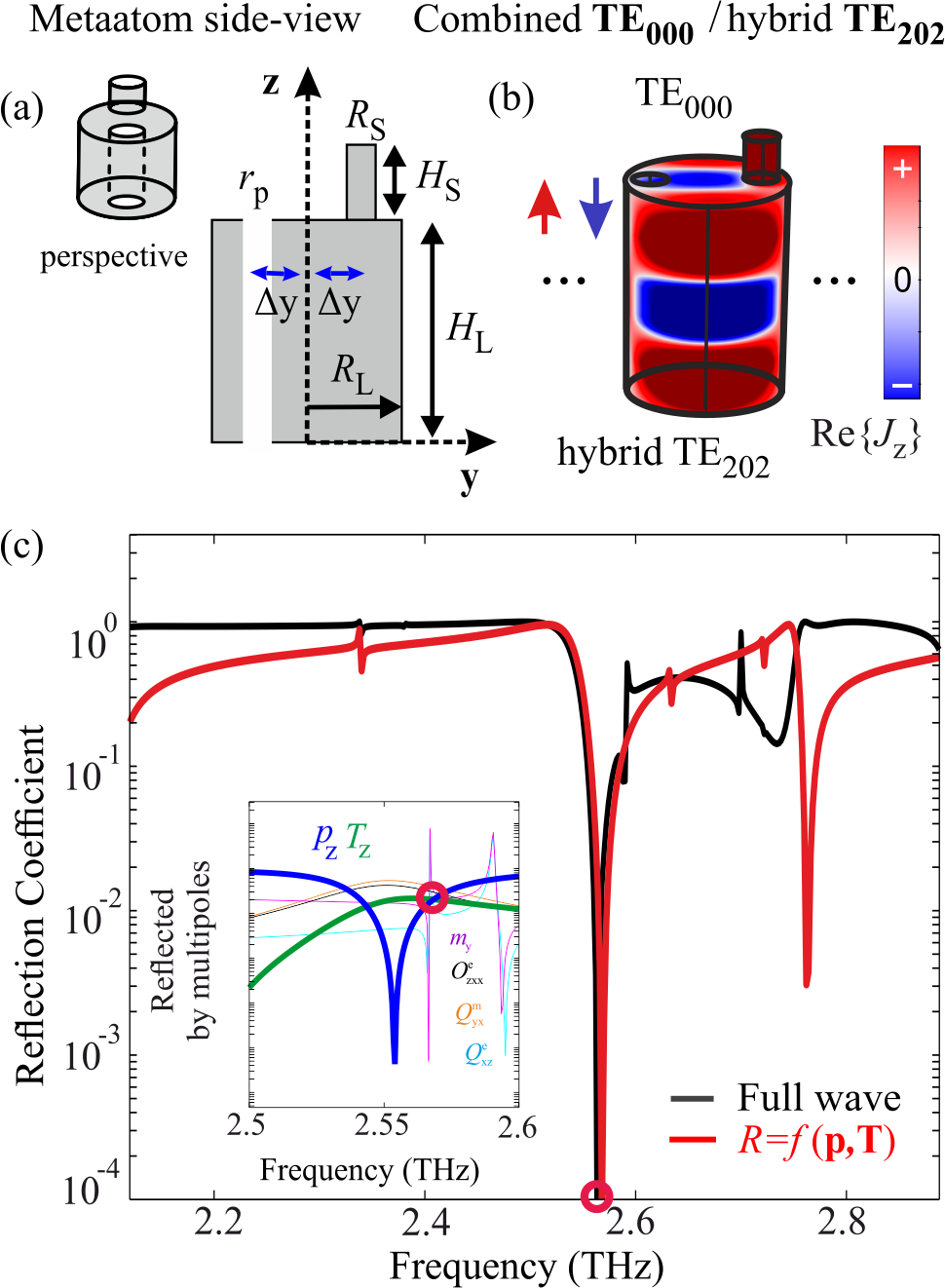}
\caption{\label{fig:perforated_reflection} (a) Schematic view of the perforated metaatom including the off-center small cross-section part ($r_\mathrm{p}=R_\mathrm{S}$). The corresponding metasurface sustains a dynamic anapole generated by the cancellation of the $p_\mathrm{z}$ and $T_\mathrm{z}$ moments; the latter originates from quadrupole Mie mode split as a result of breaking of the circular symmetry. (b) The distribution of the polarization currents of the hybrid quadrupole mode that entails, amongst else, enhanced toroidal dipole moment $T_\mathrm{z}$  (c) Reflection coefficient, assuming $E_\mathrm{z}$ polarization and normal incidence, calculated by full wave analysis (black) and as interference of the reflected amplitudes of only the $p_\mathrm{z}$ and $T_\mathrm{z}$ dipoles (red); in the inset the reflected power by several multipoles is shown.}
\end{figure}

We apply this off-center perforation to a \textit{uniform} cylinder as shown in Fig.~\ref{fig:perforated_eigen}(a). Splitings in the eigenfrequencies and changes in the current distributions are produced; the latter may be as serious as to drastically modify the character of the eigenmodes [see the TE$_{20}$ case in Fig.~\ref{fig:perforated_eigen}(a)] and their multipole components [Fig.~\ref{fig:perforated_eigen}(b)]. As mentioned before, the perforation greatly affects the magnetic type modes, causing their distinctive splitting in frequency, and has minor effect on the electric type modes. We observe this in Fig.~\ref{fig:perforated_eigen}; the perforated cylinder supports the fundamental TE$_{00}$ mode, two splitted magnetic dipole modes with $\hat{\textbf{x}}$-axis and $\hat{\textbf{y}}$-axis orientation TE$_{10}^\mathrm{x}$ and TE$_{10}^\mathrm{y}$, the fundamental mixed toroidal TE$_{01}$ mode and two splitted magnetic quadrupole modes TE$_{20}^{(1)}$ and TE$_{20}^{(2)}$. The expected radiation contributions (calculated as in Fig.~\ref{fig:rods}) presented in Fig.~\ref{fig:perforated_eigen}(b) provide further insight in the electromagnetic character of each mode. Here we choose to discuss only the above mentioned since they are the significant ones for our purposes. (For example, mode TE$_{10}^\mathrm{x}$ has a dominant $m_\mathrm{x}$ contribution that will not be excited). What is interesting is that the splitted quadrupole mode TE$_{20}^{(2)}$ has a significant $T_\mathrm{z}$ contribution with respect to the non-splitted TE$_{20}$ mode seen in Fig.~\ref{fig:rods}. This is the feature we plan to exploit in the present approach, since the features of the perforated single infinite uniform cylinder presented in Fig.~\ref{fig:perforated_eigen} survive in the perforated large part of the metaatom [see Fig.~\ref{fig:perforated_reflection}(a)], the unit cell of the sculptured cylinder. The perforation is a vertical circular hole placed at the symmetric position ($y=-10$~$\mu$m) of the small off-center cross-section part and with the same radius. Note that all the features depend on the size of the perforation which we here keep constant. The new design of the metaatom involves the  
TE$_{20}^{(2)}$ hybrid quadrupole mode of the perforated cylinder, entailing electric and toroidal dipole moments with a pure electric dipole sustained by the small-cross section part of the metaatom. The parameters are set at $H_\mathrm{L}=40$~$\mu$m and $H_\mathrm{S}=8$~$\mu$m for the cross-sectional lengths, $R_\mathrm{L}=15$~$\mu$m and $R_\mathrm{S}=r_\mathrm{p}=3$~$\mu$m for the cross-sectional radii and the radius of the hole. The cylinders are periodically arranged with the $\hat{\textbf{y}}$-axis periodicity being equal to $d=40$~$\mu$m.  Fig.~\ref{fig:perforated_reflection}(c) presents the reflection response of the metasurface and in the inset of Fig.~\ref{fig:perforated_reflection}(c) we present the power contribution of the various multipoles to the reflected power. The contribution from the toroidal dipole peaks at $\sim $2.56~THz where the electric dipole exhibits a dip. The $p_\mathrm{z}$ and $T_\mathrm{z}$ power contributions cross at two points, one of which produces a dip in the reflection, satisfying the anapole condition of Eq.~(\ref{eq:1}). This is also evident by the fact that the anapole dip almost coincides with the dip produced by the interference of only the $p_\mathrm{z}$ and $T_\mathrm{z}$ contributions, $R=f(\textbf{p},\textbf{T})$ (red), although there are at this frequency strong magnetic dipole and quadrupole and electric octupole moments [see Fig.~\ref{fig:perforated_reflection}(c), inset], which apparently cancel each other out. The quite minor discrepancy in the position of the $R=f(\textbf{p},\textbf{T})$ and the total reflection dip is $\mathrm{\Delta} f/f_0=0.1 \%$, while in this case a very small leakage in the order of 1$\%$ is recorded in the reflection (calculated by the full wave analysis). Thus, we have obtained beyond any doubt, the creation of an anapole by the destructive interference of electric and toroidal dipole amplitudes almost without interference of any other multipole. This achievement was realized by our optimized metasurface composed of sculptured cylinders periodically placed along the $\hat{\textbf{y}}$-axis; each cylinder is formed by metaatoms [see Figs \ref{fig:config1_optimized}(a) and \ref{fig:perforated_reflection}(a)] serving as unit cells; the breaking of the circular symmetry was critical in eliminating parasitic contributions detrimental to both the existence and the nature of the anapole. 

\begin{figure}
\includegraphics{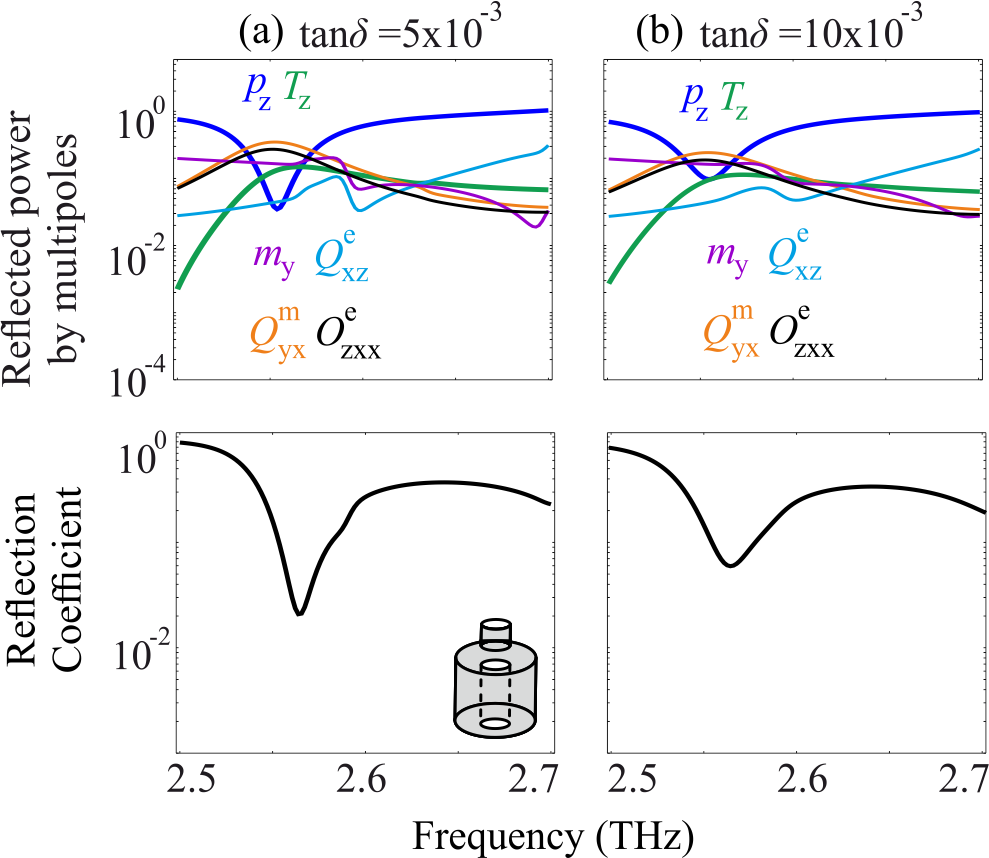}
\caption{\label{fig:losses2} Top panels: multipole contributions to the reflected power, for the optimized structure of Fig.~\ref{fig:perforated_reflection}(a); bottom panels: total reflection assuming material losses with (a) $\tan\delta =5\times 10^{-3}$ and (b) $\tan\delta=10\times 10^{-3}$. The electric dipole moment contribution increases extensively above $\tan\delta =10\times 10^{-3}$ becoming dominant, while the toroidal dipole moment contribution decreases and the crossings between the power reflected from $p_\mathrm{z}$ and $T_\mathrm{z}$ required for the anapole condition, seize to occur.}
\end{figure}

At the same time, the combination of the hybrid mixed toroidal and electric dipole modes of Fig.~\ref{fig:perforated_reflection}(a) proves to be more resilient to material losses than those of Fig.~\ref{fig:config1_optimized}(a). This is presented in Fig.~\ref{fig:losses2}, where the response of Fig.~\ref{fig:perforated_reflection}(a) metasurface around the anapole frequency range, 2.5-2.7~THz, is calculated for a material with variable loss tangent, $\tan\delta =5 \times 10^{-3}$ [Fig.~\ref{fig:losses2}(a)] and $\tan\delta =10\times 10^{-3}$ [Fig.~\ref{fig:losses2}(b)].
Top panels present the power contributions of the variable multipoles to the reflection power and bottom panels present the full wave calculated reflection. We observe that as material losses increase, the $Q$ factors of the responses decrease, leading to less sharp resonances.

Additionally, the electric dipole moment contribution increases while the toroidal dipole moment contribution decreases. The crossings between the contribution reflected from the electric dipole $p_\mathrm{z}$ and toroidal dipole $T_\mathrm{z}$ remain up to the loss tangent of $\tan\delta =10\times 10^{-3}$ [Fig.~\ref{fig:losses2}(b)]. For even greater losses, the electric dipole moment contribution becomes dominant, while the toroidal dipole moment contribution vanishes.

\newpage
\section{Conclusions}
We presented the design principle, the optimized specifications, and the simulation results of all-dielectric, polaritonic metasurfaces, consisting of sculptured periodically arranged cylinders that sustain non-radiating alternating current distributions, i.e. the so called dynamic anapole state. The sculptured cylinders are formed by the periodic repetition of a metaatom breaking the circular symmetry and serving as their building block; two different designs of the metaatom were presented based on different origins for the toroidal dipole. We show that the anapole emerges from the destructive interference of only two properly modified (as a result of the circular symmetry breaking) multipole moments: the origin of the first is mainly the TE$_{00}$ electric dipole Mie mode and the origin of the second is either the mixed toroidal dipole Mie mode TE$_{01}$ or the split magnetic quadrupole Mie mode, TE$_{20}$; the notations TE$_{00}$, TE$_{01}$, and TE$_{20}$ are those found in an infinite, uniform free-space standing cylinder. It was proven that the breaking of the circular symmetry can effectively eliminate the persistent magnetic contributions from the frequency of interest and allow thus the almost pure cancellation by destructive interference of the electric and toroidal dipole radiation leading to the anapole state. In the case that the toroidal dipole moment originates from the splitting of TE$_{20}$ mode, an additional symmetry breaking perforation in each metaatom enhances the performance with respect to the anapole realization. Both designs of the metaatom have been evaluated regarding the anapole sensitivity to material dissipation losses proving to be reasonably resilient for actual implementations. Notice that the designed metasurfaces can be realized using established fabrication techniques such as direct laser writing. 

\section{Acknowledgments}

This work was supported by the European Union's Horizon 2020 Future Emerging Technologies call (FETOPEN-RIA) under grant agreement no. 736876 (project VISORSURF), Odysseas Tsilipakos acknowledges the financial support of the Stavros Niarchos Foundation within the framework of the project ARCHERS ("Advancing Young Researchers Human Capital in Cutting Edge Technologies in the Preservation of Cultural Heritage and the Tackling of Societal Challenges"). The authors would also like to thank professor Costas M. Soukoulis for useful suggestions and valuable support.

%


\end{document}